# Spin polarized induction of quantum correlations – entanglement using a 2 MeV proton beam channeling

**Vesna Berec[1, 2]\***

[1]*Department of Physics, University of Belgrade,*
[2]*Institute of Nuclear Sciences, P. O. Box 522, 11001 Belgrade*

**J. S. Braunstein[3]**

[3]*The University of Texas at Austin, 1 University Station C1602, Austin TX 787124*

In solid-state hybrid electron-nuclear spin systems quantum entanglement plays vital role in allowing accessible transfer of information between subatomic particles, regardless of the host lattice coordination spatial geometry, revealing the powerful resource for nuclear quantum states engineering. Here we present study of 2 MeV superfocused channeled proton (SCP) beam induced polarization of atom-photon correlated states, established in isotopically purified silicon nanocrystal. Two level entangling interaction which couples an initial quantum state to two possible light–matter states via silicon nanocrystal interface is presented. The anisotropic hyperfine coupling is demonstrated by strong mixing of quantum states within the control mechanism of the coherent proton pulse sequence. Obtained results reveal the mutual predictable correlation of particles of energy/matter, by using the fully broadcastable and precise hybrid electron-nuclear spin qubit manipulations which can be exploited for the speed-superior communication channels keeping at the same time the maximum degree of data preservation.

*Corresponding author: bervesn@gmail.com

## 1. INTRODUCTION

The ability to generate, control and transfer the atom–photon quantum correlation-entanglement between light-matter interfaces [1 - 6] represents the central topic of recent developments toward the fields of quantum electrodynamics and quantum optics. However, in the presence of noise it is hard to produce, precisely asses and to classify the dynamics of quantum states according to their entanglement properties [7]. Thus, the transmission of information – quantum correlated light/matter states over large distances mainly relies on the detectors accuracy to adequately extract the detection probability from the dark count of detector intrinsic noise [8]. Accordingly, the long-term subwavelength stability which can cover the whole communication distance has proved very challenging owing to a difficulty of integrating stabilized path length fluctuation until the generation of desired entangled pairs. For that purpose, in spite of typically fragile resource of entanglement which can be easily destroyed by noise, the high fidelity transmission of information, in real experimental conditions, usually depends on the achievable generation of pure states [9]. Because the experimental preparation of such pure or singlet states is difficult as a result, the initial states are produced with variable degree of entanglement. To overcome these limitations, Werner [10] has explored entangled mixed states with regard to simple spin measurements, which appear to be nonlocal concerning the fidelity of quantum transmission. Thus, in a noisy environment the correlation between entanglement and the dissipation of purity of the two-level bipartite systems which refer to atomic and field subsystems, had recently been studied [11, 12] but without treatment or particular inclusion of the effect of the entangled mixed state.

Motivated by the fact that the mixed state quantum property as a robust entanglement resource allows the possibility of employing a noisy environment, using a recently introduced framework [13, 30, 36] we have investigated the hybrid proton mediated electron-nuclear spin dynamics in a nanocrystal channel of diamond symmetry, performing combined theoretical and experimental study. The experimental protocol for atom-photon entangled states is established via axial $\langle 100 \rangle$ nanosilicon interface. The proton spin mediation was used considering the recent high energy proton channeling studies conducted at the Super Proton Synchrotron [14] which confirmed the nonambiguous contribution of relativistic protons, confined under axial channeling regime, in the exit angular distribution of hyperchanneled particles even for maximal bending angle of silicon crystal [15].

Initially, the system containing two nuclear spins ($^1$H and $^{29}$Si nuclear spin) is coupled via anisotropic hyperfine interaction with electron spin, in a ground state. This state is further laser-excited to a metastabile triplet. After the initial preparation, the system provides all requirements for exploration of the hybrid nuclear-electron [16, 17] spin manipulations which are further mediated by a proton spin in conjunction with hyperfine-transient electron spin (via dynamic nuclear polarization [18]), using a combination of 2 MeV energy polarized channeled proton beam pulses driven in picosecond scale, and a laser excitation at a 221.7 nm [19]. The 2 MeV energy establishes an acceleration of the spin interactions due to nuclear spins long coherence lifetimes [20] and compensates weak polarization of the nuclear spins, thus overcomes the spin decoherence limit imposed by the noisy environment. The preparation of the initial state considers the hyperfine coupling induced by the proton beam pulses in a 92 nm thick nanosilicon crystal target following the experimental conditions introduced in [21].

## 2. INTERACTION MECHANISM AND METHODS

We shall now briefly elaborate the model and implemented technique. The swift 2 MeV energy protons are confined under the axial channeling regime (when the channeled proton trajectory corresponds to oscillatory motion) and strongly localized between adjacent atomic rows. We have used the axial channeling configuration in order to increase the confinement effect over ion trajectories, allowing them to be hyperchanneled, i.e., efficiently captured within one single channel [29]. In particular, we have used the 92 nm thick silicon nanocrystal to capture proton oscillatory motion between four neighboring atomic rows according to diamond lattice fcc symmetry of Si $\langle 100 \rangle$ channel, as shown in figure 1. The incident proton beam is tilted relative to z-axis, i.e., $\langle 100 \rangle$ low index axis of Si nanocrystal, for the specific values of angles below critical angle for channeling [29, 30]. The gap between the two Si lattice sites represents nanocrystal's cavity [31]. Four nanocrystal atomic planes which perpendicularly intersect the corresponding atomic rows of $\langle 100 \rangle$ channel, interact with CP field [30] producing a gap smaller than the half of the planar oscillation wavelength of the proton beam (its coherence length).

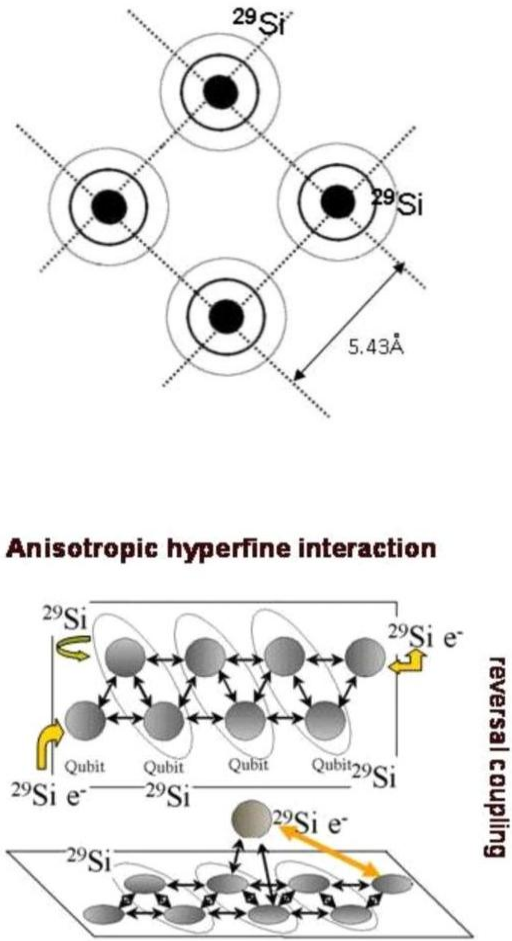

**Fig. 1.** Top: $\langle 100 \rangle$ representation of axial channel formed by the four $^{29}$Si atoms, where transversal thickness of the target corresponds to one atomic layer, a schematic view. Bottom: The hyperfine interaction mechanism, a schematic representation. The anisotropic hyperfine coupling yields the synchronization between selective rotations of the adjacent nuclear $^{29}$Si spins affected by the circularly polarized electron spin.

Nanocrystal planes are capable to act as a mirror that inverses the transverse motion and deflects the ion trajectories [21], forming the resonant cavity conditions similar to an X rays resonator [31, 32]. Based on theoretical study [33] that a high efficient ''mirror'' for charged particles can be generated by an ultrathin crystal which is tilted relative to the direction of the incident beam for typical angles smaller than $\psi_c = 6.09$ mrad, recent experiment and simulation study [21] exposed a method for deflection of channeled ions, guided via coherent interactions, in silicon crystal of thickness parameter smaller than 100 nm. More recently, axial confinement produced by ultrathin silicon ion channeling, was experimentally confirmed for nonrelativistic protons, focused through a 55 nm thick [001] Si membrane [34]. It was shown that the transverse phase space can be populated by hyperchanneled ion trajectories, allowing an effective resource of transversely polarized particles.

Basically, we are considering here a structure comprising two main Hamiltonian components: the ion-atom confinement potential, acting inside the silicon nanocrystal cavity, and the internal-spin-Hamiltonian. In order to describe the system properties and dynamics of the continuum ion-atom interaction potential in the nanocrystal, we include the Hamiltonian governing the oscillatory motion of ions

$$H = (1/2)m\left(p_\perp^2 + U(r)\right) = E\left(\psi_x^2 + \psi_y^2\right) + U(r),$$
$$E_\perp = E\psi^2 + U(r) \qquad (1)$$

where $E_\perp$ and $p_\perp$ denote ion transverse energy and momentum, $\psi_x$ and $\psi_y$ are $x$ and $y$ components of scattering small angle with respect to the low index channel axis. The proton trajectories are obtained in the Molière approximation of the Thomas-Fermi interaction potential [29, 30]

$$U_i(r) = \frac{2Z_1 Z_2 e^2}{d} \sum_{i=1}^{3} \alpha_i K_0\left(\beta_i \frac{r}{a}\right), \qquad (2)$$

where $Z_1$ and $Z_2$ are the atomic numbers of the proton and the atom, respectively, $e$ is the electron charge, $d$ is the quantum displacement from the harmonic oscillator ground state, $r$ is the distance between the proton and atomic strings, $a_0$ is the Bohr radius, $a = \left[9\pi^2/128 Z_2\right]^{\frac{1}{3}} \cdot a_0$ is the atom screening radius, and $K_0$ is the zero order modified Bessel function of the second kind with the fitting parameters:
$(\alpha_i) = (0.35, 0.55, 0.10)$, $(\beta_i) = (0.30, 1.20, 6.00)$ [29].

The internal-spin-Hamiltonian comprises: nuclear and electron spin qubit, localized in nanosilicon target, as depicted in figure 1. and a mediator spin system of channelled protons [30, 35]. The Hamiltonian for the case system placed in an external magnetic field $B_x$, is

$$H = \omega_e S_z + \omega_{^1H} I_z^{^1H} + \omega_{^{29}Si} I_z^{^{29}Si} + S_z \otimes \left[\sum_{n \in ^{29}Si, ^1H} \left[A_n I_z^n + B_n I_x^n\right]\right]. \qquad (3)$$

The Hamiltonian includes electron spin component $S_z$ along the direction $\hat{z}$ of the static external field $\mathbf{B}_0$. The operator of the nuclear spins $I^n$ refers to: hyperchanneled protons ($^1H$), and the $^{29}Si$, nuclei

(nanocrystal target is 99% isotopically purified). $\omega_e$, $\omega_H$ and $\omega_S$ are Zeeman frequencies for electron, $^1H$, and $^{29}Si$ nuclei, respectively. $A_n$ and $B_n$ are coefficients of the hyperfine coupling. Such system possesses the primary orientation dependence from the coefficients of the hyperfine interaction.

The nuclear spins, i.e., $^1H$ and the $^{29}Si$, are affected by the hyperfine anisotropic term $BS_zI_x$ which couples longitudinal component of the electron spin with the transverse component of the nuclear spin. The $B_x$ field is applied perpendicular to $\hat{z}$ axis which corresponds to the excitation laser propagation. For the atom–lattice experiment we use the CP field stored in the collective lattice mode to manipulate charged particles at ideal conditions for separate qubit control. The optically generated states in that sense afterword can be transferred as wave-packets along the transmission line for the multiple-qubit sessions.

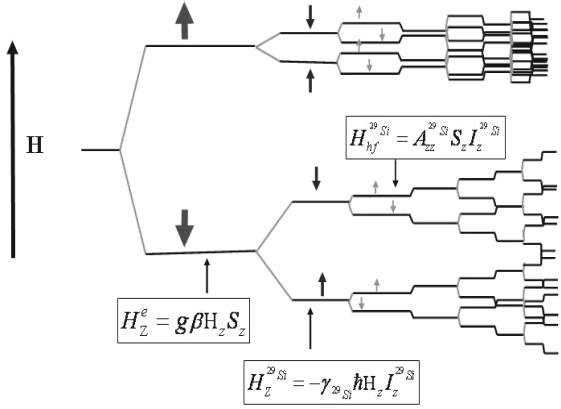

**Fig. 2**: $^{29}Si$ energy level structure corresponding to $B_x = 2$ T, shows the Zeeman splitting for electron and $^{29}Si$ nuclei and hyperfine splitting for $^{29}Si$ nuclei between $3s^2\ 3p^2\ 3P_2$ and $3s\ 3p^3\ 3D_3^0$ level, a graphic representation. The electron spin Zeeman term is given by $H_Z^e = g\beta H_z S_z$. $g$ and $\beta$ are electron g-tensor and the Bohr magneton, respectively. The nuclear spin Zeeman term is $H_z^{^{29}Si} = -\gamma_{^{29}Si}\hbar H_z I_z^{^{29}Si}$; $\gamma_{^{29}Si} = -5.31$ is the gyromagnetic ratio for the $^{29}Si$ target and $H_{hf}^{^{29}Si} = A_{zz}^{^{29}Si} S_z I_z^{^{29}Si}$ is the hyperfine electron-nuclear spin interaction. $A_{zz}$ is the hyperfine second rank tensor along the z axis, where the hyperfine constant is: $A = -160.1 \pm 1.3$ MHz.

## 3. DISCUSSION AND RESULTS

The quantum transition between the two nodes which represent the coupled system of two lattice atoms and the silicon nanocrystal cavity-lattice mode $|E_c\rangle = \frac{1}{\sqrt{2}}(|e,0\rangle + |g,1\rangle)$ in a form of the resulting quantum state *1-2-C* is given by the following relation:

$$|T\rangle = \frac{1}{2}\{|e_1\rangle(|i_2\rangle + |g_2\rangle)|0\rangle + |g_1\rangle(|i_2\rangle - |g_2\rangle)|1\rangle\}. \quad (4)$$

Here $|T\rangle$ refers to three particle entangled state of $\frac{1}{2}$ spins. The CP field coupled with $^{29}Si$ atoms generates oscillations between the ground $|e,0\rangle$, excited state $|g,1\rangle$ and triplet state $|s,1\rangle$. As a result, the atom-cavity quantum state decay/produce horizontally polarized photons denoted by 1 and vertically polarized photons or 0.

It is possible to entangle a stream of such qubits by transmitting them through a thin silicon nanocrystal, exposed to a high collimated channeled proton beam in order to change the polarization, after the state of the qubits can be read off on exit plane of the crystal. During a read out measurement, a qubit collapse into a 0 or 1 state. Because of the primary orientation-dependable configuration of the triplet states, in addition to polarized beam exposition, tilting of a target additionally allows for the precise control of a crystal orientation with selection of the specific electron triplet polarization [37]. The laser pulse is used synchronously to a proton beam in this protocol in order to drive a well defined superposition of the Zeeman states, given by eq. (3) for $B_x = 2$ T condition, between electron and nuclear spins $|\uparrow\rangle$, $|\downarrow\rangle$, toward the excitation (S-$T_0$ qubit basis [38]) which represents a mixture of the two spin states: the electron $|T_0\rangle \equiv |\uparrow\downarrow\rangle + |\downarrow\uparrow\rangle/\sqrt{2}$ and the nuclear $|S\rangle \equiv |\uparrow\downarrow\rangle - |\downarrow\uparrow\rangle/\sqrt{2}$ spin state.

Atom-photon entanglement is produced by forming a state with multiple decay channels. The system excitation in one of the $^3P$ states is induced via laser (P = 50 nW) imprinting the effective spin coupling states onto a photon thus allowing decay through the $P$-$D^0$ channel via photon $|V\rangle$, polarized parallel to the electron-nuclear quantization axis, and photon $|H\rangle$, polarized perpendicularly to electron-nuclear quantization axis.

A short proton pulses thus provide a direct selective spin control by flipping the state of the excitation selectively in a picosecond time scale ~ 13ps length. Figure 3 (a) denotes the ground state of the mediator $|g\rangle$ and the first excited state $|e\rangle$. The nuclear $^{29}Si$ spin and electron spin are coupled to the excited state of the mediator (channelled $^1H$ spin) via anisotropic Heisenberg interaction given by eq. (3). A transition from the system's singlet to triplet state is provided by selected $\pi$ pulse rotation relative to quantization z axis (represented on Bloch sphere) through the angle $\varphi = J(\varepsilon)/\hbar \leq \psi_c$. $J(\varepsilon)$ is the exchange coupling between different energy $\varepsilon$ sublevels.

A proton beam than controls the precession over $\vartheta$-angle by the ellipticity which is equivalent to the spin polarization relative to low index $\langle 100\rangle$ axis of silicon nanocrystal.

The one-dimensional thermal vibration amplitude of the nanocrystal's atoms is 0.0074 nm [29, 30, 37] and the average frequency of transverse motion of protons moving close to the channel axis is equal to $5.94\times10^{13}$ Hz.

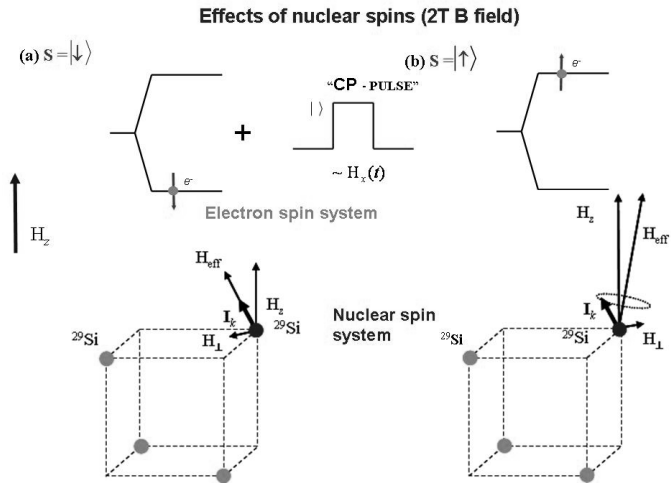

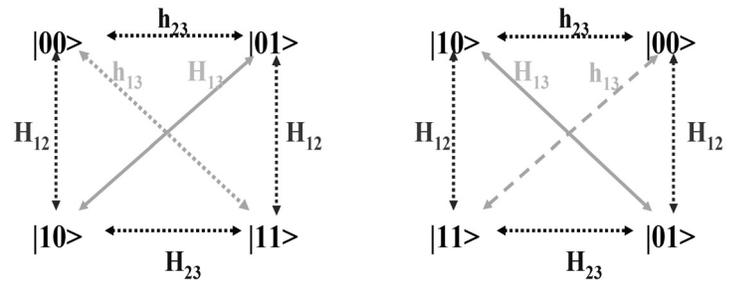

**Fig. 4**: Exchange interaction - pictorial presentation. Symmetric exchange interaction (indicated by grey full line) $H_{ij} = J(s_{x,i}s_{x,j} + s_{y,i}s_{y,j}) + K(s_{x,i}s_{y,j} - s_{y,i}s_{x,j})$ couples entangled $^{29}Si$ nuclear qubit states |01> and |10> (left). The antisymmetric exchange interaction (indicated by grey dotted line) $h_{ij} = j(s_{x,i}s_{x,j} - s_{y,i}s_{y,j}) + k(s_{x,i}s_{y,j} + s_{y,i}s_{x,j})$ couples the states |00> and |11> (right).

**Fig. 3**: Schematic of electron spin transitions between the two energy sublevels induced by CP pulse. Top: the probability of a "spin-flip" between the states $m_s$ = +1 and $m_s$ = -1 is maximized when the CP pulse has energy equal to $\hbar\omega_0$. Bottom: at initial condition, in equilibrium, the nuclear spin qubits are aligned in the z direction of $B_o$. When a CP pulse is applied at the resonant frequency, the nuclear spins start to be affected by the $B_x$ field and precess about effective $B_x$ direction. By controlling the angle and duration of a CP pulse, the magnetization is rotated into the x-y plane.

The majorization protocol is performed over two two-level subsystems referring to singlet–triplet, s,t basis which includes the coherent mixtures of basis states that form an equal incoherent mixture of the four Bell's entangled states, see eq. (6), governed by the exchange Hamiltonian, as represented in figure 4:

$$|00\rangle_t \equiv \langle 0|_A \langle 0|_B, \quad |11\rangle_t \equiv \langle 1|_A \langle 1|_B$$
$$|10\rangle_{s,t} \equiv \langle 1|_A \langle 0|_B, \quad |01\rangle_{s,t} \equiv \langle 0|_A \langle 1|_B, \quad (5)$$

$$|\Phi_\pm\rangle = \frac{1}{\sqrt{2}}(|0\rangle_A|1\rangle_B \pm |1\rangle_A|0\rangle_B)$$
$$|\Psi_\pm\rangle = \frac{1}{\sqrt{2}}(|0\rangle_A|0\rangle_B \pm |1\rangle_A|1\rangle_B). \quad (6)$$

Here, the qubit representation:
$\langle m_S m_I | = |0\rangle|0\rangle, |0\rangle|1\rangle, |1\rangle|0\rangle, |1\rangle|1\rangle$ corresponds to a Zeeman product states $\langle m_S m_I | = |\uparrow\uparrow\rangle, |\uparrow\downarrow\rangle, |\downarrow\uparrow\rangle, |\downarrow\downarrow\rangle$; the electron/nuclear spin states are denoted as: $\uparrow = 1/2$, $\downarrow = -1/2$.

We have used the protocol which simultaneously increases both purity and entanglement, at the cost of decreasing the ensemble size of initial photon pairs. In addition, we have theoretically explored the region of maximally entangled Werner state corresponding to generated photon-ion entangled systems close to the limit of maximally entangled state.

Obtained entangled Werner state is given by by:

$$\hat{\rho}_w = (1-p)\frac{1}{4}\hat{1}_4 + p|\Phi_\pm\rangle\langle\Phi_\pm|, \quad (7)$$

where $|\Phi_\pm\rangle = \frac{1}{\sqrt{2}}(|HH\rangle \pm e^{i\varphi}|VV\rangle)$, and $\hat{1}_4$ denotes the identity matrix. Here $|HH\rangle$ denotes two horizontally polarized and $|VV\rangle$ represents two vertically polarized photons, while $\varphi$ corresponds to the proton beam incident angle [30, 35]. We have performed superoperator tomography of the density matrix states (16×16) to analyze the probabilities for the concurrence which refers to Werner entangled states, figure 5. (a), (b), (c).

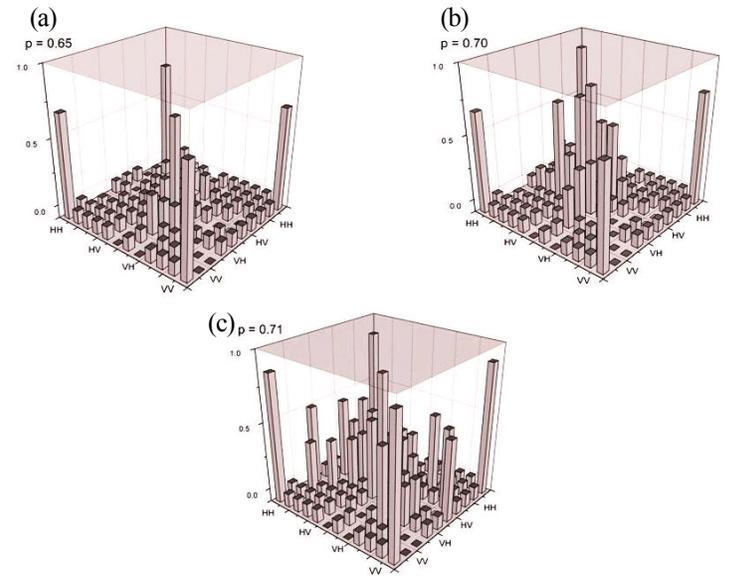

**Fig. 5:** Tomography superoperator results for the real components of Werner states (10 × 10 matrix) for probability range corresponding to $p \leq 1/\sqrt{2}$, the residual 4×4 matrix imaginary components are omitted. Plots: (a) p = 0.65, (b) p = 0.70, (c) p = 0.71 correspond to the ion beam incident angle of $\varphi = 0.1\psi_c$, where $\psi_c = [2Z_1Z_2e^2/(dE)]^{1/2} = 6.09$ mrad. Silicon nanocrystal is tilted along the $\theta_x$ axis corresponding to the limit p > 1/3 which defines the maximal quantity of entanglement.

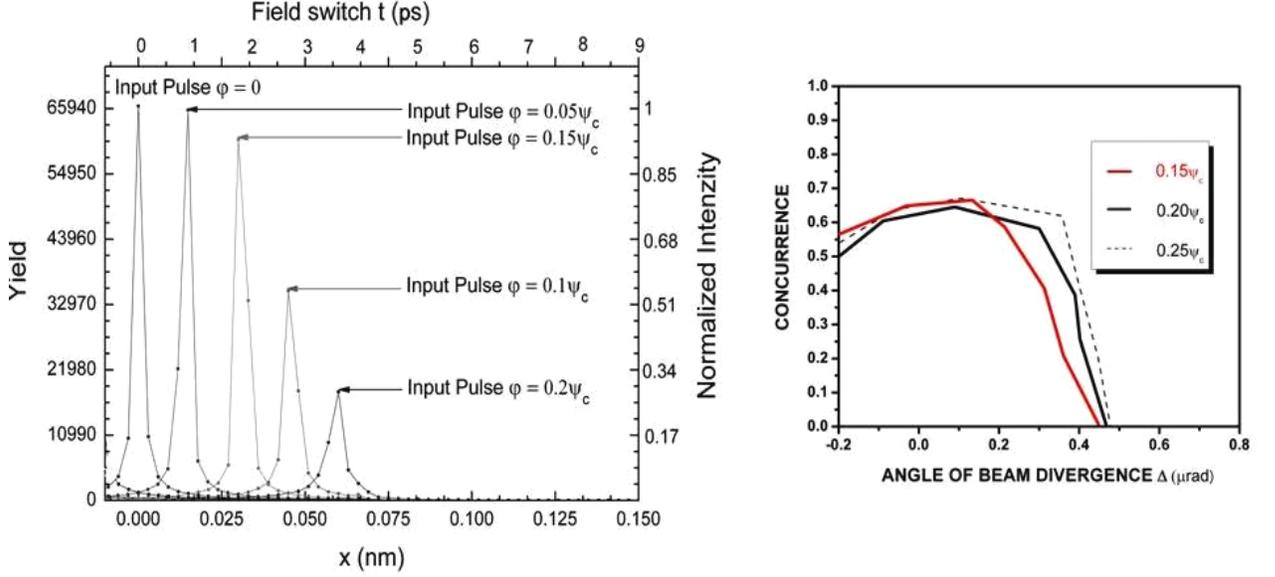

**Fig. 6** Distribution of amplitudes obtained from 2 MeV proton beam source with a 92 nm $^{29}$Si target. Left: general pulse sequence is determined in picosecond time scale for the spin polarization (responsible for rotation of the spins around the *z* axis relative to low index $\langle 100 \rangle$ axis of the silicon target) associated with the tilt angles: $\varphi = 0.05\psi_c$, $\varphi = 0.1\psi_c$, $\varphi = 0.15\psi_c$ and $\varphi = 0.20\psi_c$, in the transverse phase plane. The proton beam focusing spot covers the area in the vicinity of the low index $\langle 100 \rangle$ nanocrystal axis and allows the control for two-qubit entanglement with inhibition of the new spin polarizations for the nuclear $^{29}$Si spins in adjacent sites.

Right: Dependences of the measure of entanglement quantity – concurrence as a function of the angle of the beam divergence Δ, for the specific tilt angles of the proton beam, numerical results.

### 3.1 Experimental protocol

The entanglement properties of nuclear $^{29}$Si qubits which belong to two nanosilicon sites are established by exposing the system to synchronized Ti: sapphire ultrafast laser at 82 MHz repetition frequency, in conjunction to 2 MeV energy channeled proton beam pulses, tilted relative to quantization axis by angles: $\varphi = 0.05\psi_c$, $\varphi = 0.1\psi_c$, $\varphi = 0.15\psi_c$ and $\varphi = 0.20\psi_c$, as presented in figure 6 (left). The excitation pulses are chosen to match the limits of the Bohr radius, denoted with peak at 20% of the critical angle for channeling $\psi_c$ (relative to tensor principal axis). The anisotropic hyperfine coupling between the triplet state of ($^1$H mediated) polarized electron spin and the $^{29}$Si nuclear spin is further coherently manipulated via selective proton beam pulses where the spins precessing is utilized under a modified Meiboom–Gillecho sequence [38, 39]. The pulses are generated periodically; picosecond phase interval is indicated in figure 6 (left). Note that at the end of the each sequence the sign of the pulses reverses in order to compensate the accumulation of the phase noise. Entangled photon pairs are generated via spontaneous Stokes Raman transition at the corresponding nanocrystal site with different polarization states $|H\rangle$ and $|V\rangle$, where the final states are Werner entangled, $|\Psi\rangle_\pm = \frac{1}{\sqrt{2}}(|H\rangle \pm e^{i\varphi}|V\rangle)$. The atom (site) excitation is provided with femtosecond $\sigma^+$ polarized pulse, resulting in spontaneous emission to either state l or state 0, in addition to emission of a photon in $|H\rangle$ or $|V\rangle$ polarization state. By using the polarization beam splitter [40] it is possible then to reflect the photons with polarization $|V\rangle$ $|-\rangle$ and transmit the polarization $|H\rangle$ $|+\rangle$ photons.

The probe laser and atomic qubits are initially in coherent states: $|\psi_1\rangle$ and $|\psi_2\rangle$ respectively, which result in the unnormalized two level entangled state between two photon states and two atomic states as:

$|\Phi\rangle = \frac{1}{2}[\psi_{2+}|\psi_1\rangle \otimes |\psi_2\rangle_+ - i\psi_{2-}|-\psi_1\rangle \otimes |\psi_2\rangle_-]$ with $|\psi_2\rangle_\pm$ denoting the even and odd coherent superposition states:

$$|\psi_2\rangle_\pm = \frac{1}{\psi_{2\pm}}(|\psi_2\rangle \pm |-\psi_2\rangle),$$

$$\psi_{2\pm} = \sqrt{2(1 \pm e^{-2|\psi_2|^2})}. \qquad (8)$$

The corresponding concurrence, in figure 6(left), which refers to the presence of entanglement, is obtained corresponding to [7, 22] as:

$$C = \sqrt{\left[1 - \exp(-4|\psi_1|^2)\right]\left[1 - \exp(-4|\psi_2|^2)\right]}. \qquad (9)$$

To confirm these two entangling operations, we put the two spins into a Werner mixed state and perform the effective spin density matrix tomography [41]. It can be seen that the amount of entanglement (entanglement of formation) shows rapid increase only close to upper limit of the Werner probability range. Observed effect strongly depends on the purified distance between the Werner and maximally entangled mixed state.

This means that one can manipulate and precisely control the quantum entanglement of state of eq. (8) by varying the CP and **B** field intensity, and the angle of incidence of the proton beam relative to nanocrystal low index $\langle 100 \rangle$ axis, as presented in figure 6. Thus, the stronger the intensity of the proton beam (corresponding to tilt angles ~ 10-25% of the critical angle for channeling), for a given 2T field, the entanglement of the quantum state (eq. (7)) is stronger. To coherently drive the quantum transition between $S \leftrightarrow T_0 \leftrightarrow T_\pm$ nuclear states, in a pulsed regime, laser with 20 μeV bandwidth is used to generate approximately 150 fs pulse width, much less than the corresponding singlet–triplet splitting.

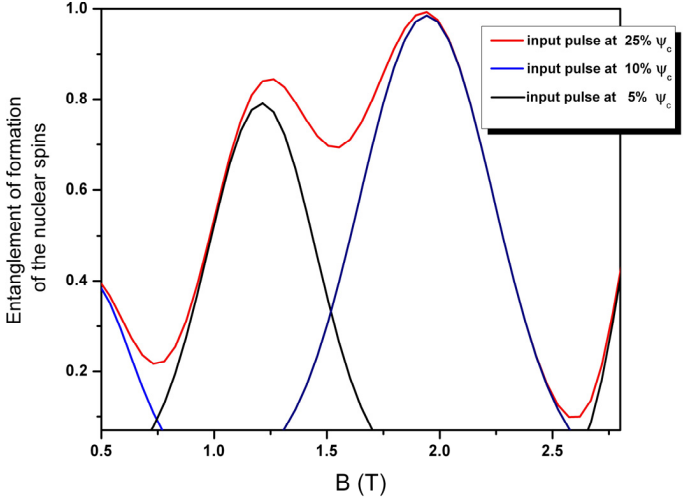

**Fig. 7**: Numerically obtained dependence of the quantity measure of established mixed state entanglement for nuclear spins - entanglement of formation [23] as a function of the implemented magnetic field, considering various channeled proton beam angles.

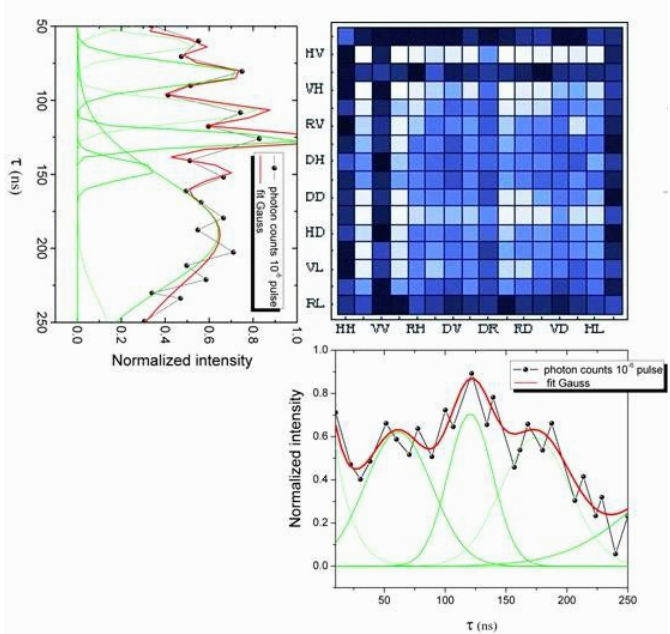

**Fig. 8**: Quantum Monte-Carlo simulation of two photon count-down emission probability as a read out of $m_S = 0$ and $m_S = \pm 1$ states, representing a function of the time interval, τ, during a pulse sequence in x-y plane responsible for the corresponding hyperfine transition. Probabilities are obtained using a superoperator density matrix tomography for polarization states:

$$\begin{array}{cccc}
|H\rangle_1 \otimes |H\rangle_2 & |R\rangle_1 \otimes |H\rangle_2 & |D\rangle_1 \otimes |R\rangle_2 & |V\rangle_1 \otimes |D\rangle_2 \\
|H\rangle_1 \otimes |V\rangle_2 & |R\rangle_1 \otimes |V\rangle_2 & |D\rangle_1 \otimes |D\rangle_2 & |V\rangle_1 \otimes |L\rangle_2 \\
|V\rangle_1 \otimes |V\rangle_2 & |D\rangle_1 \otimes |V\rangle_2 & |R\rangle_1 \otimes |D\rangle_2 & |H\rangle_1 \otimes |L\rangle_2 \\
|V\rangle_1 \otimes |H\rangle_2 & |D\rangle_1 \otimes |H\rangle_2 & |H\rangle_1 \otimes |D\rangle_2 & |R\rangle_1 \otimes |L\rangle_2 .
\end{array} \quad (10)$$

A full $16 \times 16$ Hermitian matrix, obtained from the superoperator tomography, is further included in quantum Monte Carlo simulation. It represents a linear combination of the different coincidence measurements, where:

$$|D\rangle = \frac{\langle H| + \langle V|}{\sqrt{2}}, \ |L\rangle = \frac{\langle H| + i\langle V|}{\sqrt{2}}, \ \text{and} \ |R\rangle = \frac{\langle H| - i\langle V|}{\sqrt{2}}.$$

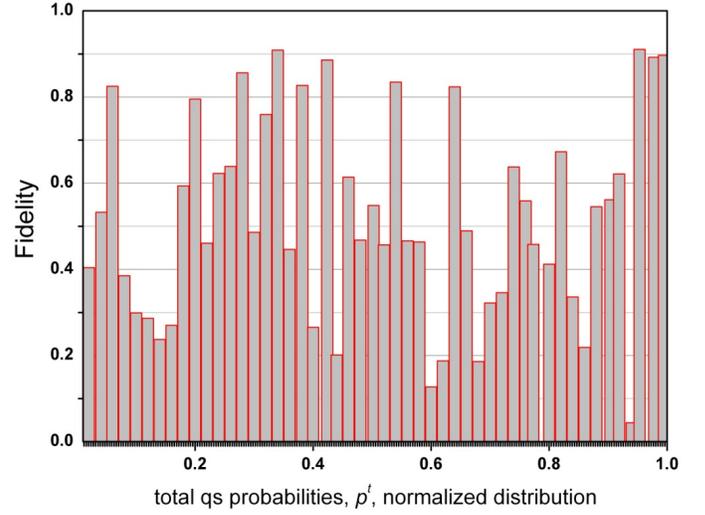

**Fig. 9**: The Uhlmann-fidelity corresponding to mean values and standard deviations obtained from data shots of 10000 quantum Monte-Carlo iterations each using the quantum state reconstruction (eq. (8)).

Figure 9 represents the fidelity, $F$, of an obtained set of state tomography data, expressed by

$$F(\psi_{2\pm}) = \sum_\rho p^t F(\rho, \psi_{2\pm}). \quad (11)$$

The corresponding discrepancies, $\Delta F$, with respect to a desired state $\psi_{2\pm}$ (see eq. 8) are obtained as

$$\Delta F(\psi_{2\pm}) = \left( \sum_\rho p^t \left( F(\rho, \psi_{2\pm}) - F(\psi_{2\pm}) \right)^2 \right)^{\frac{1}{2}}. \quad (12)$$

Considering $N$ iterations to detect $f_i$ results from probability distribution function given by

$$f(p, c, N) = \prod_j \frac{p \cdot D_p\left(c_j, m_k\right) + (1-p) \cdot D_p\left(c_j, m_l\right)}{D_p\left(c_j, m_k\right) + D_p\left(c_j, m_l\right)}, \quad (13)$$

where, $D_p\left(c_j, m_{\{i,j\}}\right) = \frac{m_{\{i,j\}}^{c_j} e^{-m_{\{i,j\}}}}{c_j!}$ represents the normalized probabilities of detecting $c_j$ counts for a given Poisson distribution centered around mean values $m_{\{k,l\}}$ for $m_S = 0$ and $m_S = \pm 1$ states, one can obtain the binomial distribution of a probability distribution function given in eq. (13). Here the total probability, $p^t$, of a quantum state $\psi_{2\pm}$ in the measurement process is given by the product of all probabilities as

$$p^t = \frac{\Pi_i f\left(tr(P_i\rho), n_i, N\right)}{\sum_\rho \Pi_i f\left(tr(P_i\rho), n_i, N\right)}. \quad (14)$$

In this way for each iteration one can utilize series of deterministic measurements which are non-destructive [42] with respect to the joint atom-photon state.

Finally, in order to asses the quantum transfer which amounts to a perfect single-use quantum channel, the corresponding communication between obtained entangled states is topologically described by the entanglement of percolation threshold [43, 44] between two arbitrary spatial centers, where a partially entangled state can be converted into a maximally entangled state with a certain probability that depends on the initial amount of entanglement. The probability for realization of entangled states, $\rho_{(AB)}$, between spins of two-level subsystems $A$ and $B$ at positions $i$ and $j$ is given by

$$P|\psi_1\rangle = \sum_{i,j}\left(p_{i,j}|00\rangle_t + \sqrt{1-p_{i,j}^2}|11\rangle_t\right)$$
$$P|\psi_2\rangle = \sum_{i,j}\left(p_{i,j}|10\rangle_{s,t} + \sqrt{1-p_{i,j}^2}|01\rangle_{s,t}\right). \quad (15)$$

This means that every quantum operation or $q$-swap between N bipartite settings can be implemented with probability $P_q$, $1-P_q$. Assignment of the values for each $P_q$ specifies the quantum transmission. For large communication lines/networks, conversion probability to singlet states can be assigned to $P^2$ and it is independent of the distance between the nodes and of the size of the network [6, 42]. In particular, two arbitrary nodes can be connected by a singlet if both belong to the same cluster. With the excitation probability $\varepsilon \approx L_0/L$, where the maximal distance between nodes is $L_0$, the time needed in the entanglement connection process can be estimated to $T \propto (L/L_0)^2$ and scales polynomially with the communication distance. The transmitting success probability is on the order of $O(\varepsilon^2\eta^2 e^{-L_0/L_{att}})$ by considering the channel attenuation, where $\eta$ denote the detection efficiency. The time needed in this process is $T \approx T_{cc}/\varepsilon^2\eta^2 e^{-L_0/L_{att}}$, with $T_{cc} = L_0/c$ which denotes the communication time.

## 4. SUMMARY

We have demonstrated generation and precise control of channelled proton beam induced atom-photon quantum correlations for QED-based entanglement and described conversion protocol which efficiently maps the ion-atom into photon quantum state, allowing the efficient correlation of entangled state between arbitrary localized spatial centers. Coherent control is utilized through precession in the proton exchange field which is initialized via 2 MeV energy superfocused proton beam pulses. The noise reduction through the correlation process is achieved by establishing the specific circumstances when two quantum objects - spin qubits form a unique mixed quantum state in the composite system, type: singlet - triplet. In that context, the process of coupling of electron with ½ nuclear quantum spin states in silicon nanocrystal target, mediated by the polarized nuclear spin states of channeled protons through the quantum entanglement, allows the transfer of information originally deposited in the electrons to the spin state of the host $^{29}$Si. The result is an extremely fast transfer of quantum information in long-lived quantum state (polarization) of a nuclear spin, further addressable to a photon, with corresponding polarization/frequency. Obtained results support further investigation toward ion-beam manipulation of mixed entangled states, emphasizing the correlated quantum state transfer of a new type of particle /field interaction under the MeV energy channelling regime.

## Acknowledgement
The authors thank to I. Anicin on fruitful discussions during the early stage of this work. This work is supported by the Ministry of Science and Education, Republic of Serbia.